\documentclass[journal=rsm]{CUP-JNL-DTM}%

\RequirePackage[style=nejm,autocite=superscript,url=false]{biblatex}
\AtEveryBibitem{%
  \clearfield{note}%
  \clearfield{urldate}
}
\usepackage{graphicx}
\usepackage{multicol,multirow}
\usepackage{amsmath,amssymb,amsfonts}
\usepackage{mathrsfs, rsfso}
\usepackage{amsthm}
\usepackage{rotating}
\usepackage{appendix}
\usepackage{ifpdf}
\usepackage[T1]{fontenc}
\usepackage{newtxtext}
\usepackage{newtxmath}
\usepackage{textcomp}
\usepackage{xcolor}
\usepackage{lipsum}
\usepackage[colorlinks,allcolors=blue]{hyperref}
\usepackage{mdframed}

\theoremstyle{definition}

\numberwithin{equation}{section}

\usepackage[frozencache, cachedir=minted]{minted}
\usepackage[none]{hyphenat} 

\usepackage{array,multirow,ragged2e,hhline}

\usepackage{rotating}   
\usepackage{pdflscape}

\usepackage{soul} 
\soulregister\autocite7

\jname{}
\articletype{Software Focus}

\setstcolor{violet}

\begin{document}

\begin{Frontmatter}

\title[Article Title]{Meta-analysis with the glmmTMB R package}

\author[1,2]{Coralie Williams\textsuperscript{*}}
\author[3]{Maeve McGillycuddy}
\author[4]{Mollie Brooks}
\author[5,6]{Benjamin M. Bolker}
\author[7]{Ayumi Mizuno}
\author[1,8]{Yefeng Yang\textsuperscript{*}}
\author[9]{Wolfgang Viechtbauer}
\author[1,2]{David I. Warton}
\author[1,7]{Shinichi Nakagawa\textsuperscript{*}}

\address[1]{\orgdiv{Evolution \& Ecology Research Centre}, \orgname{UNSW}, \orgaddress{\city{Sydney}, \postcode{2052}, \state{NSW},  \country{Australia}}}

\address[2]{\orgdiv{School of Mathematics and Statistics}, \orgname{UNSW}, \orgaddress{\city{Sydney}, \postcode{2052}, \state{NSW},  \country{Australia}}}

\address[3]{\orgdiv{Stats Central, Mark Wainwright Analytical Centre}, \orgname{UNSW}, \orgaddress{\city{Sydney}, \postcode{2052}, \state{NSW},  \country{Australia}}}

\address[4]{\orgdiv{National Institute of Aquatic Resources}, \orgname{Danish Technical University}, \orgaddress{\city{Kongens Lyngby}, \postcode{2800}, \country{Denmark}}}

\address[5]{\orgdiv{Department of Mathematics \& Statistics}, \orgname{McMaster University}, \orgaddress{\city{Hamilton},\state{Ontario}, \country{Canada}}}

\address[6]{\orgdiv{Department of Biology}, \orgname{McMaster University}, \orgaddress{\city{Hamilton}, \state{Ontario},  \country{Canada}}}

\address[7]{\orgdiv{Department of Biological Sciences, Faculty of Science}, \orgname{The University of Alberta}, \orgaddress{\city{Edmonton}, \state{Alberta},  \country{Canada}}}

\address[8]{\orgdiv{Department of Biosystems Engineering}, \orgname{Zhejiang University}, \orgaddress{\city{Hangzhou}, \state{Zhejiang},  \country{China}}} 

\address[9]{\orgdiv{Department of Psychiatry and Neuropsychology}, \orgname{Maastricht University}, \orgaddress{\city{Maastricht},  \country{The Netherlands}}}

\address[*]{
\email{coraliewilliams@outlook.com} 
}


\keywords{multilevel models, hierarchical models, open-source, evidence synthesis, random effect structure, model-based inference}

\abstract{
Two common formulations of meta-analytical models include the standard two-stage normal--normal models, which synthesise estimated effect sizes, and the one-stage generalised linear mixed model (GLMM), which directly model the underlying outcome data across studies.
The general-purpose \texttt{glmmTMB} R package provides flexible response distributions and random-effect covariance structures through Template Model Builder (TMB). Its existing functionality can fit one-stage meta-analytic GLMM specifications.
However, incorporating known sampling variances and covariances in the conventional two-stage inverse-variance formulation of meta-analysis was previously not easily accomplished in \texttt{glmmTMB}. 
Here, we introduce \texttt{equalto}, a new covariance structure in \texttt{glmmTMB} that allows users to supply a known sampling error variance–covariance matrix when fitting meta-analytic models. This enables explicit modelling of heteroscedasticity and dependence among sampling errors. 
Using simulations, we show that \texttt{glmmTMB} produces estimates identical to those from the corresponding \texttt{metafor} package functions for normal--normal models and similar estimates for GLMM specifications. We illustrate these models using published meta-analysis datasets in medicine, evolutionary ecology, and the social sciences.
With the addition of the \texttt{equalto} covariance structure, \texttt{glmmTMB} now provides a unified and flexible framework for fitting two-stage normal--normal models and one-stage meta-analytic GLMMs, including multivariate specifications. These model classes can be fitted using the same \texttt{glmmTMB()} function, expanding the R toolkit available for evidence synthesis.
}

\end{Frontmatter}

\newpage
\begin{mdframed}
\section*{Highlights}

\subsubsection*{What is already known}
\begin{itemize}
    \item Meta-analysis is commonly implemented either by synthesising estimated effect sizes in a two-stage normal--normal inverse-variance model or by fitting the underlying study outcomes directly in a one-stage GLMM.
    \item The general-purpose mixed model R package \texttt{glmmTMB} supports a wide range of distribution families, random-effect covariance structures, and separate dispersion and zero-inflation formulas, with fast estimation of complex GLMMs.
\end{itemize}

\subsubsection*{What is new}
\begin{itemize}
    \item We introduce \texttt{equalto}, a new covariance structure in the \texttt{glmmTMB} R package that enables conventional two-stage inverse-variance meta-analysis.
    \item The new implementation allows users to specify known variance-covariance matrices for sampling errors and complements the existing \texttt{glmmTMB} functionality for fitting one-stage meta-analytic GLMMs.
\end{itemize}

\subsubsection*{Potential impact for \textit{Research Synthesis Methods} readers}
\begin{itemize}
    \item With the addition of \texttt{equalto}, \texttt{glmmTMB} provides a unified and flexible meta-analytical toolkit for fitting two-stage normal--normal models and one-stage meta-analytic GLMMs through a single model-fitting function and an interface familiar to users of the \texttt{lme4} R package.
    \item Meta-analytical models fitted in \texttt{glmmTMB} can draw on its broader modelling capabilities, including complex and multivariate random-effect covariance structures, including temporal, spatial and phylogenetic terms, and, where appropriate, dispersion and zero-inflation formulas.
\end{itemize}
\bigskip
\end{mdframed}

\textbf{Research Synthesis Keywords:}  R package, software, mixed-effects models, dependent effect sizes.



\section{Introduction}


Meta-analysis provides a quantitative synthesis of results across studies, allowing more general conclusions from a broader evidence base\autocite{gurevitch_metaanalysis_2018, glass_primary_1976, lau_summing_1998}. 
A common meta-analytic model is a linear mixed model, where variation across studies and effect sizes are modelled as random effects, and known sampling errors are incorporated. 
Several software tools support meta-analysis, for example software purpose-built for meta-analysis RevMan \autocite{thecochranecollaboration_review_}, OpenMEE \autocite{wallace_openmee_2017}, MetaXL \autocite{elmakaty_mastering_2025}, and more general statistical software such as STATA, Jamovi\autocite{jamovi_jamovi_}, JASP \autocite{jaspteam_jasp_2025}, and R\autocite{r-core-team_language_2022}. 

R statistical software holds multiple dedicated packages for fitting meta-analytic models supported by its versatility, open-source, and active development\autocite{polanin_review_2017}. In particular, the \texttt{metafor} R package\autocite{viechtbauer_conducting_2010} is widely used (2.1 million downloads as of 2025), providing extensive functionality for both standard and advanced meta-analytic models. For binary, proportion, and count outcome data, meta-analytic models can be formulated as generalised linear mixed models (GLMMs) and can be fitted with \texttt{metafor} using the \texttt{rma.glmm()} function. More broadly, because meta-analysis can be expressed as a special case of a mixed model, it can also be fitted using general-purpose software such as packages for generalized linear mixed models (GLMMs), which may provide additional flexibility for complex data structures and non-Gaussian outcomes.


General-purpose packages for fitting GLMMs, such as \texttt{lme4}\autocite{bates_fitting_2015} in R, are long established and widely used across many disciplines, offering a flexible framework with an interface (e.g. formula syntax) that is straightforward for researchers to apply.
An emerging R package, \texttt{glmmTMB}\autocite{brooks_glmmtmb_2017}, has become an alternative broad-purpose GLMM package in R, with over 1.8 million downloads since its inception and around 61,000 downloads per month at the time of writing. Model estimation in \texttt{glmmTMB} uses Template Model Builder (TMB), an R package named for its use of C++ templates to define model likelihoods. `TMB' supports efficient maximum likelihood estimation by integrating Gaussian random effects using the Laplace approximation and computing gradients using automatic differentiation\autocite{kristensen_tmb_2016}.
However, users do not need to interact directly with TMB or write C++ code, as \texttt{glmmTMB} makes this estimation framework available through a familiar interface similar to \texttt{lme4}\autocite{bates_fitting_2015, brooks_glmmtmb_2017}.
\texttt{glmmTMB} not only provides efficient and fast estimation but also allows the specification of a wide variety of distributions, covariance structures \autocite{kristensen_covariance_2025}, as well as dispersion and zero-inflated modelling. Recent developments have further enhanced its scope, such as an implementation for reduced-rank analysis\autocite{mcgillycuddy_parsimoniously_2025}, broadening modelling possibilities. 
With its computational efficiency, diverse modelling options, and user-friendly interface, \texttt{glmmTMB} is well-suited for fitting meta-analytic models, offering a broadly applicable framework that accommodates diverse data types and study designs beyond the scope of specialised meta-analytical software.

In this paper, we introduce a new covariance structure in \texttt{glmmTMB}, called \texttt{equalto}, which allows users to specify a random effect covariance structure that is exactly equal to a user-specified sampling variance-covariance matrix. This addition complements the existing one-stage GLMM functionality in \texttt{glmmTMB}, allowing conventional two-stage and one-stage meta-analytical models to be fitted using the same function. More broadly, \texttt{glmmTMB} supports model components not currently available in dedicated meta-analysis packages, including multiple random effects in dispersion models, zero-inflation structures, complex covariance structures, and non-linear smooth functions.

In what follows, we describe meta-analytic models under the traditional inverse-variance framework and the GLMM framework, then describe \texttt{glmmTMB} function functionalities, including model specification and results summary. As a proof of concept, we present the results of a simulation study comparing estimates from the \texttt{glmmTMB} function with the widely used \texttt{metafor} package for both the one-stage and two-stage models. Finally, we demonstrate the application of the new implementation with published meta-analysis datasets in medicine, social sciences, and evolutionary ecology.

\section{Methods}

Effect sizes, the outcome variable in meta-analysis, are derived from raw data or summary statistics of primary studies.
Broadly, effect sizes can be classified in three categories: (1) single descriptor measures such as proportions or rates, (2) comparative measures that quantify differences between groups or conditions, and (3) measures of association such as correlation coefficients or regression slopes\autocite{white_choice_2020}. The appropriate effect-size measure for single or comparative analyses depends on the underlying data type (e.g.\, continuous, binary, count). 

In the following sections, we first describe `conventional' meta-analytical models within a linear mixed model framework, and then show how this framework can be extended to generalized linear mixed models (GLMMs) to model non-Gaussian primary study data.

\subsection{Meta-analytic models} \label{section2.1}

Consider $k$ independent effect size estimates, indexed by $i=1,\dots,k$. Each observed effect size $y_i$ is assumed to estimate an underlying true (unknown) effect $\theta_i$ with sampling error $e_i$:
\begin{equation}
y_i = \theta_i + e_i, \qquad e_i \sim \mathcal{N}(0, v_i),
\label{eq1}
\end{equation}
where $v_i$ is the known sampling variance. Thus, conditional on $\theta_i$, the observed effects $y_i$ are assumed to be normally distributed with mean $\theta_i$ and variance $v_i$.  

To allow for variation in the true effects across studies, we assume that each $\theta_i$ deviates randomly from an overall mean effect $\mu$:
\begin{equation}
\theta_i = \mu + u_i, \qquad u_i \sim \mathcal{N}(0, \tau^2),
\label{eq2}
\end{equation}
where $\tau^2$ represents the between-study variance (heterogeneity). 
Equations~\ref{eq1} and \ref{eq2} define the \textit{Normal–Normal model} in meta-analysis, in which both the observed study estimates $y_i$ and the study-specific true effects $\theta_i$ are assumed to follow (approximately) normal distributions, with $\theta_i \sim \mathcal{N}(\mu,\tau^2)$.
This Normal-Normal model forms the basis for most conventional random-effects meta-analytic methods, but represents a convenient modelling formulation rather than a strict requirement of random-effects meta-analysis. It is referred to as the "two-stage" approach as it involves two steps: (1) using individual data points to derive effect sizes and sampling variances; (2) fitting the estimated effect sizes and sampling variances in a meta-analytical model. These assumptions may not always hold for all effect-size measures or underlying data-generating processes. \\ 

In many applications, effect sizes are not independent but clustered, for example, when multiple effect sizes contribute to the same study. To account for such dependencies among effect sizes, additional random effects can be introduced, leading to a multilevel meta-analysis model (also called hierarchical effects (HE) model\autocite{pustejovsky_metaanalysis_2022, hedges_robust_2010}). When effect sizes are nested within studies, let $y_{ij}$ be the $i$th effect size from study $j$, then we have the following model:
\begin{equation}
y_{ij} = \mu + u_j + m_{ij} + e_{ij},
\label{eq3}
\end{equation}
where $u_j \sim \mathcal{N}(0,\tau^2)$ represents a between-study random effect with variance $\tau^2$, $m_{ij} \sim \mathcal{N}(0,\sigma^2)$ captures within-study heterogeneity among multiple effect sizes, and $e_{ij} \sim \mathcal{N}(0, v_{ij})$ is the known sampling error with variance $v_{ij}$, which is derived or calculated from primary study results. 

In cases with multiple effect sizes per study, the effect sizes may be derived from the same sample or subject (e.g.\ shared controls or multiple outcome measures), inducing dependence among sampling errors. To account for this, we can model the dependence by assuming a correlation among effect sizes within studies:

\begin{equation}
e_{ij} \sim \mathcal{N}(0, V)
\label{eq4}
\end{equation}
where $V$ is a block-diagonal variance--covariance matrix. The $i$th block contains the sampling variances of the effect sizes for study $i$ on the diagonal, and the corresponding covariances on the off-diagonals, assuming a level of correlation among effect sizes based on the information available from the primary studies. 
Equations \ref{eq3} and \ref{eq4} together form the Correlated and Hierarchical effects (CHE) model described in Pustejovsky and Tipton, 2022\autocite{pustejovsky_metaanalysis_2022}.


To explore heterogeneity, Equation \ref{eq3} can be extended to include moderators (i.e.\ explanatory or fixed effects) that may explain between-effect variation and adjust for confounding factors, which are often study or measurement characteristics.

\subsection{Meta-analysis within a generalized linear mixed-effects model (GLMM) framework} \label{section2.2}

The meta-analysis framework in Section~\ref{section2.1} can be parameterised to model raw binary or count data from primary studies using generalized linear mixed models (GLMMs). GLMMs link the linear predictor to an appropriate response distribution (e.g.\, binomial, Poisson), allowing non-Gaussian outcomes to be modelled directly without the need to derive standardised effect size estimates. This approach, often referred to as the “one-stage” analysis in the literature (vs the "two-stage" analysis in Section \ref{section2.1})\autocite{papadimitropoulou_onestage_2019, riley_twostage_2023, kontopantelis_comparison_2018}, provides a unified framework for meta-analysis across a wide range of outcome types and study designs enabling the use of raw data and avoiding transformation of effect sizes prior to modelling \autocite{jackson_comparison_2018, lin_metaanalysis_2020, stijnen_random_2010, simmonds_general_2016}.

More generally, GLMMs consist of three components:  
(i) a \textit{distributional component} specifying the distribution family of the response variable,  
(ii) a \textit{linear predictor} given by $\eta_i$, and  
(iii) a \textit{link function} $g(\cdot)$ that relates the expected value of the response to the linear predictor. Given these components, a GLMM can be specified as:
\begin{equation}
g(\mu_i) = \eta_i = \mathbf{x}_i^{\mathsf{T}}\boldsymbol{\beta} + \mathbf{z}_i^{\mathsf{T}}\mathbf{b},
\end{equation}
where $\mu_i = \mathbb{E}(y_i)$, $\mathbf{x}_i$ and $\mathbf{z}_i$ are design row-vectors for fixed and random effects (rows of the respective design matrices), $\boldsymbol{\beta}$ is a vector of fixed-effect parameters, and $\mathbf{b}$ is a random-effects vector where $\mathbf{b} \sim \mathcal{N}(\mathbf{0}, \boldsymbol{\Sigma})$. Conditional on $\mathbf{b}$, the observed outcomes $y_i$ are assumed independent and to follow a distribution $\mathcal{F}$, such that $y_i \mid \mathbf{b} \sim \mathcal{F}(\mu_i, \phi)$, where $\phi$ denotes a dispersion or residual variance parameter.
In meta-analytic applications without moderators, $\boldsymbol{\beta}$ reduces to a single overall mean effect $\beta$, and $\boldsymbol{\Sigma}$ collapses to the between-study variance $\tau^2$ (i.e.\, $\mathbf{b}\sim\mathcal{N}(0,\tau^2)$). 
In the Supplementary Material, we provide examples of GLMM formulations for meta-analyses of binary data (Binomial–Normal models) and count data (Poisson–Normal models), i.e.\ the so-called “one-stage” approach. GLMMs can be estimated using likelihood-based approximation methods (e.g., Laplace approximation or adaptive Gauss–Hermite quadrature) which provide tractable approximations to the marginal likelihood of the random effects \autocite{bolker_generalized_2009}. In \texttt{glmmTMB}, models are estimated using the Laplace approximation implemented via automatic differentiation \autocite{brooks_glmmtmb_2017}.

\section{Software functionality}

The \texttt{glmmTMB} R package supports one-stage meta-analytic GLMMs with its existing functionality and, through the new \texttt{equalto()} covariance structure, conventional two-stage inverse-variance models. Below, we first outline the main components and modelling capabilities of \texttt{glmmTMB()}, then demonstrate how to fit two-stage models using \texttt{equalto()}.

\subsection{\texttt{glmmTMB} function features}
The \mintinline{r}{glmmTMB} function provides a flexible framework for fitting GLMMs suitable for meta-analysis\autocite{brooks_glmmtmb_2017}. Table~\ref{tab:glmmTMB-functionalities} summarises four components of the \texttt{glmmTMB}\autocite{brooks_glmmtmb_2017} function relevant to meta-analytic modelling: (i) a flexible formula interface, with the same syntax as \texttt{lme4}\autocite{bates_fitting_2015}, allowing the specification of multiple fixed and random effects with different covariance structures\autocite{kristensen_covariance_2025} ; (ii) a range of distribution families and link functions; (iii) a dispersion model formula (\verb|dispformula|) for modelling the scale parameter; (iv) a zero-inflation model formula (\verb|ziformula|) for modelling excess or structural zeros in meta-analysis of count data. Together, these features support both conventional meta-analytic models (Section~\ref{section2.1}) and GLMM formulations (Section~\ref{section2.2}).

\subsection{Data and model specification}

To demonstrate the new functionality for incorporating known sampling variances, we fit the conventional Normal–Normal meta-analysis model described in Section~\ref{section2.1}. 

To begin, we load the \texttt{glmmTMB} package:
\begin{minted}{R}
# load library
library(glmmTMB)
\end{minted}

We will use a published meta-analysis dataset \texttt{dat.assink2016}\autocite{assink_risk_2015} available in the \texttt{metadat} package\autocite{viechtbauer_metadat_2025}. The dataset includes 17 studies comparing recidivism between delinquent juveniles with and without a mental health disorder. Effect sizes are provided in terms of standardized mean differences (\texttt{yi}), with positive values indicating a higher prevalence of recidivism in the group of juveniles with a mental health disorder. Corresponding (known) sampling variances (\texttt{vi}) are also provided. 

\begin{minted}{R}
library(metadat)
dat <- dat.assink2016 |>
  transform(id = as.factor(id), #set row indexes as factor for equalto()
            g = 1)
head(dat,8)
# study esid id      yi     vi pubstatus year deltype g
#     1    1  1  0.9066 0.0740         1  4.5 general 1
#     1    2  2  0.4295 0.0398         1  4.5 general 1
#     1    3  3  0.2679 0.0481         1  4.5 general 1
#     1    4  4  0.2078 0.0239         1  4.5 general 1
#     1    5  5  0.0526 0.0331         1  4.5 general 1
#     1    6  6 -0.0507 0.0886         1  4.5 general 1
#     2    1  7  0.5117 0.0115         1  1.5 general 1
#     2    2  8  0.4738 0.0076         1  1.5 general 1
\end{minted}

Assuming independence among sampling errors, we construct the variance-covariance matrix $\mathbf V$, with the provided sampling variances \texttt{vi} ($v_i$ in Equation \ref{eq1}) along the diagonal:

\begin{minted}{R}
V <- diag(dat$vi)
round(V[1:5,1:5],3) 
#       [,1] [,2]  [,3]  [,4]  [,5]
# [1,] 0.074 0.00 0.000 0.000 0.000
# [2,] 0.000 0.04 0.000 0.000 0.000
# [3,] 0.000 0.00 0.048 0.000 0.000
# [4,] 0.000 0.00 0.000 0.024 0.000
# [5,] 0.000 0.00 0.000 0.000 0.033 
\end{minted}

\bigskip

We can then fit the \textbf{random-effects meta-analysis} specified in Equations \ref{eq1} and \ref{eq2} by fitting a Gaussian mixed-model in \texttt{glmmTMB}. The model includes two random effect terms: one corresponding to the between-study random effect and the other representing the sampling error:

\begin{minted}{R}
fit.rm <- glmmTMB(yi ~ 1 + (1|study) + equalto(0 + id|g,V),
                       dispformula=~0,
                       data=dat,
                       REML=TRUE)
\end{minted}

The known sampling error variances are incorporated within the \mintinline{r}{equalto()} structure, which has three components: 

\begin{enumerate}
    \item \mintinline{r}{0 + id} indexes the rows of the effect size and their corresponding sampling variance. The variable \mintinline{r}{id} must be a factor variable.
    \item \mintinline{r}{g} is a grouping factor whose levels correspond to independent random‐effects vectors with shared covariance parameters. In meta-analysis, it is typically specified as a single level for all observations, so that a single random‐effect is fitted to the entire dataset. If \mintinline{r}{g} has $K$ levels, this implies $\mathrm{Var}(\mathbf b)=\mathbf I_{K}\otimes\mathbf V$ where $\mathbf V$ is fixed.
    \item $\mathbf V$ is a user-supplied symmetric sampling error variance–covariance matrix, with variances on the diagonal and optional covariances on the off-diagonal. 
\end{enumerate}


We set \texttt{dispformula=$\sim$0} to fix the residual variance to approximately zero (implemented in \texttt{glmmTMB} as \texttt{\detokenize{sqrt(.Machine$double.eps)}}, a very small positive value, rather than exact zero as it is not a valid finite value under the Gaussian dispersion parameterisation), such that no residual variance is estimated and the remaining variability is absorbed by the random effects.

The model uses a restricted maximum likelihood (REML) as it is generally preferable for Gaussian responses; note that \texttt{glmmTMB} defaults to maximum-likelihood (ML), so you need to set \texttt{REML=TRUE} explicitly. In the \texttt{metafor} package, for instance, REML is used as the default for the \texttt{rma}, \texttt{rma.uni}, and \texttt{rma.mv} functions.

It is possible to fit the same model that assumes no correlation among sampling errors without the \texttt{equalto()} structure, using the \texttt{map} and \texttt{start} arguments in \texttt{glmmTMB}:

\begin{minted}{R}
fit.rm2 <- glmmTMB(yi ~ 1 + (1|study),
                   dispformula = ~0 + id,
                   map = list(betadisp = factor(rep(NA, nrow(dat)))),
                   start = list(betadisp = log(sqrt(dat$vi))),
                   data = dat,
                   REML = TRUE)
\end{minted}

\bigskip

The models above assume independence among sampling errors. This assumption may not hold in this dataset because there are multiple effect-size estimates per study and some may be derived from the same subject.
To model dependence among sampling errors (Equation \ref{eq4}), we construct an approximate sampling error variance–covariance matrix assuming a constant within-study correlation of approximately $\rho=0.6$ using the \mintinline{r}{vcalc()} function from \texttt{metafor}:

\begin{minted}{R}
library(metafor)
VCV <- vcalc(vi=vi, cluster=study, obs=id, data=dat, rho=0.6)
VCV[1:5,1:5]
#       [,1]  [,2]  [,3]  [,4]  [,5]
# [1,] 0.074 0.033 0.036 0.025 0.030
# [2,] 0.033 0.040 0.026 0.019 0.022
# [3,] 0.036 0.026 0.048 0.020 0.024
# [4,] 0.025 0.019 0.020 0.024 0.017
# [5,] 0.030 0.022 0.024 0.017 0.033
\end{minted}

The \mintinline{r}{vcalc()} function constructs a variance–covariance matrix using user-supplied sampling variances (\mintinline{r}{vi}), accounting for dependence among multiple observations (\mintinline{r}{obs = id}) within clusters (\mintinline{r}{cluster = study}) by assuming a constant within-cluster correlation (\mintinline{r}{rho = 0.6}).

\bigskip

To fit the \textbf{multilevel meta-analysis} specified in Equation \ref{eq3}, which accounts for dependence among effect sizes ($y_ij$), we add the within-study component $m_{ij}$ via an additional random effect term \texttt{(1|id)} and use the approximate sampling error VCV (Equation \ref{eq4}) derived above:
\begin{minted}{R}
fit.ml <- glmmTMB(yi ~ 1 + (1|study) + equalto(0 + id|g,VCV) + (1|id),
                  dispformula = ~0,
                  data = dat,
                  REML = TRUE)
\end{minted}

This model is equivalent to fitting a model where we allow the residual term to be modelled (removing the \texttt{dispformula=$\sim$0}) which will represent the within-study component $m_{ij}$:

\begin{minted}{R}
fit.equalto <- glmmTMB(yi ~ 1 + (1|study) + equalto(0 + id|g,VCV),
                      data = dat,
                      REML = TRUE)
\end{minted}

\subsection{Summarising model results}

\begin{figure}[p] 
   \centering
   \includegraphics[width=1\linewidth]{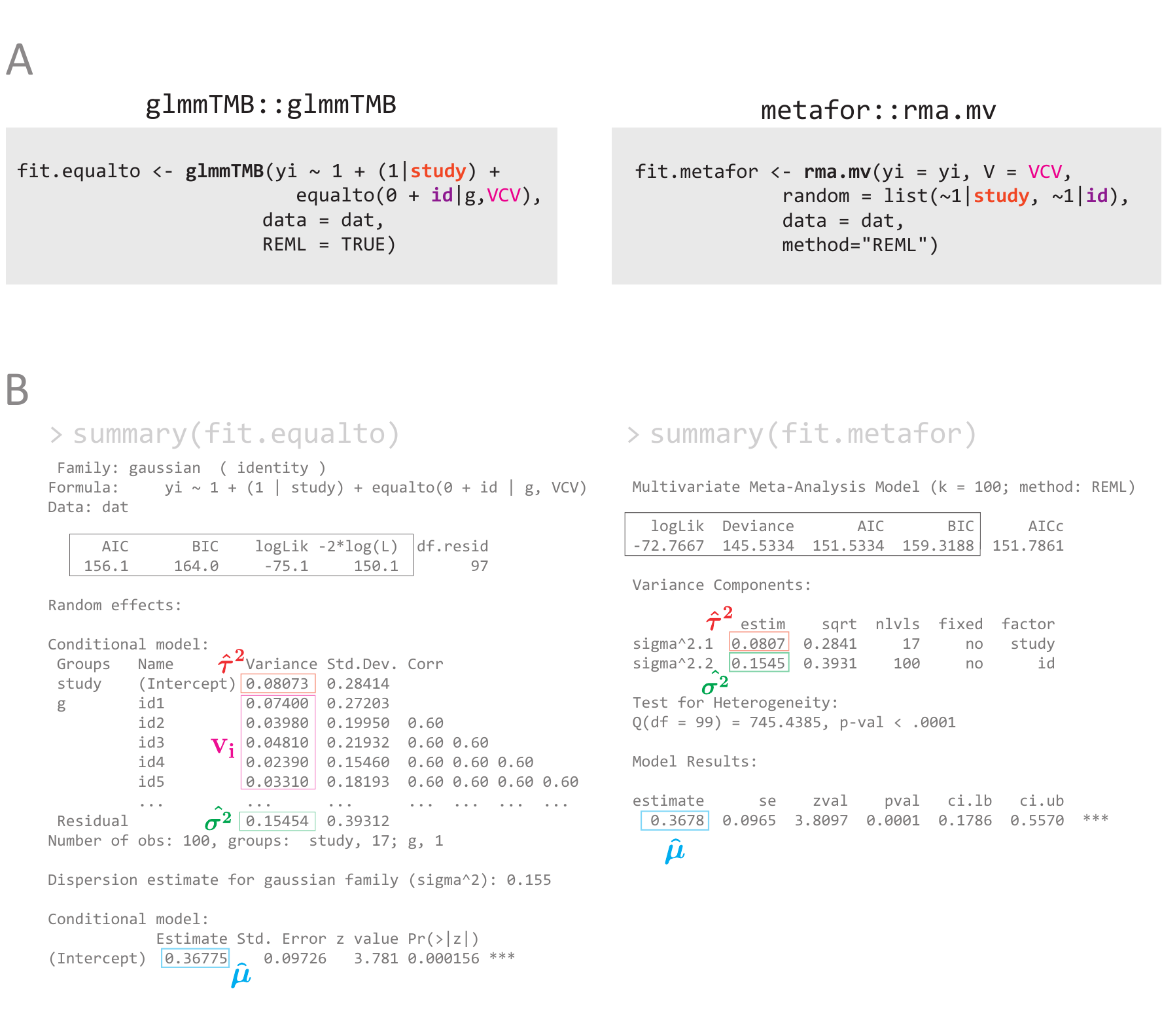}
   \caption{Model specification and output of a multilevel meta-analysis (Equation \ref{eq3}) with \texttt{glmmTMB} using \texttt{equalto} (left side) and the \texttt{rma.mv} function from the \texttt{metafor} package (right side). \\
   \textbf{A}. Multilevel meta-analysis specification with one study-level random effect (\texttt{study}) and a within-study variance component (\texttt{id}) modelled as the residual variance for the \texttt{glmmTMB} model. The sampling error variance component (\texttt{id}) with a fixed variance-covariance matrix specified as \texttt{VCV} is specified via the \texttt{equalto} covariance structure in \texttt{glmmTMB} function and via the \texttt{V} argument in the \texttt{rma.mv} function.  \\
   \textbf{B}. Model outputs using the function \texttt{summary()} which shows identical values across the two functions for the overall mean estimate $\hat\mu$, the among study variance component estimate $\hat\tau^2$, and the within-study variance component estimate $\hat\sigma^2$}
   \label{fig:fig1}
\end{figure}

Model outputs can be extracted and summarised following model fitting. Panel \textbf{A} of Figure~\ref{fig:fig1} presents the multilevel meta-analysis model specification using the \texttt{glmmTMB} and \texttt{metafor} R packages. Panel \textbf{B} of Figure~\ref{fig:fig1} presents the results of calling \texttt{summary()} on fitted models from \texttt{glmmTMB} and \texttt{rma.mv} function from \texttt{metafor}. Wald and profile confidence intervals for \texttt{glmmTMB} models can be obtained using the \mintinline{r}{confint()} function. Note that the log-likelihood, and therefore, the information criteria and deviance, differ slightly between the summary output of the two implementations (but the model estimates are identical). To obtain the same likelihood in \texttt{metafor}, supply the argument \mintinline{r}{control = list(REMLf = FALSE)} to \texttt{rma.mv}.

Reporting the overall mean and variance components of the conditional model is often of interest in meta-analyses. Table~\ref{tab:mlm-summary} provides code to extract fixed-effect estimates and variance components from the multilevel model in \texttt{glmmTMB} showcased in Figure~\ref{fig:fig1} alongside the equivalent code for the \texttt{rma.mv} function. If moderators are included, the code will return a vector of moderator coefficients in the order they were included in the conditional model formula. If additional random effects are present, their variance component estimate can be extracted by referencing the corresponding random-effect (grouping-factor) name.

\begin{landscape}
\begin{table}[p]
\centering
\caption{Components of the \texttt{glmmTMB} function for fitting meta-analysis}
\label{tab:glmmTMB-functionalities}
\setlength{\tabcolsep}{6pt}
{\fontsize{9}{11}\selectfont
\begin{tabular}{|>{\RaggedRight\arraybackslash}m{0.14\textwidth}|
                >{\RaggedRight\arraybackslash}m{0.14\textwidth}|
                >{\RaggedRight\arraybackslash}m{0.33\textwidth}|
                >{\RaggedRight\arraybackslash}m{0.42\textwidth}|
                >{\RaggedRight\arraybackslash}m{0.31\textwidth}|}
\hline
\multicolumn{2}{|c|}{\textbf{Component}} & \textbf{Application} & \textbf{Formula syntax / argument} & \textbf{Reference} \tabularnewline
\hhline{|=====|}

\multirow[c]{10}{=}{\textbf{Conditional model formula}}%
& \textbf{Fixed and random effects}
& Fit multiple fixed and random effects
& e.g. \verb! y ~ 1 + x + (1|g)!
& Same flexible formula syntax as the \texttt{lme4} package \autocite{bates_fitting_2015}. \tabularnewline
\hhline{|~----|}

& \multirow[c]{8}{=}{\textbf{Covariance structures for random effects}}
& Fixed covariance structure (to specify sampling variances and their correlations)
& \verb|equalto|
& \multirow[c]{8}{*}{\href{https://cran.r-project.org/web/packages/glmmTMB/vignettes/covstruct.html}{Covariance structures vignette}} \tabularnewline
\hhline{|~|~|-|-|~|}

&  & Independence and compound-symmetry structures
& \verb|diag|, \verb|cs|, \verb|homcs|
&  \tabularnewline
\hhline{|~|~|-|-|~|}

&  & For temporal random effects
& \verb|ar1|, \verb|hetar1|, \verb|toep|, \verb|homtoep|
&  \tabularnewline
\hhline{|~|~|-|-|~|}

&  & For spatial random effects
& \verb|ou|, \verb|exp|, \verb|gau|, \verb|mat|
&  \tabularnewline
\hhline{|~|~|-|-|~|}

&  & For proportional random effects  (e.g., phylogenetic)
& \verb|propto|
&  \tabularnewline
\hhline{|~|~|-|-|~|}

&  & Reduced-rank for high-dimensional data
& \verb|rr|
&  \tabularnewline
\hhline{|=====|}

\multicolumn{2}{|l|}{\textbf{Distributional families}}
& For GLMM formulations e.g.\ meta-analyses with binary or count data ("one-stage" analysis).
& \verb|family| e.g.\verb| family = binomial(link = "logit")|
& Use \mintinline{r}{?family} and \mintinline{r}{?family_glmmTMB} to see supported families \tabularnewline
\hhline{|=====|}

\multicolumn{2}{|l|}{\textbf{Dispersion model formula}}
& Specify a one-sided formula for dispersion using fixed and/or random effects (on log link scale).
& \verb|dispformula| e.g.\verb! dispformula = ~ x + (1|g)!
& \tabularnewline
\hhline{|=====|}

\multicolumn{2}{|l|}{\textbf{Zero-inflation model formula}}
& For meta-analyses of count data, allow excess zeros (on the logit link scale).
& \verb|ziformula| e.g.\verb! ziformula = ~ x + (1|g)!
& \tabularnewline
\hhline{|=====|}


\end{tabular}
}
\end{table}
\end{landscape}


\begin{landscape}
\vfill
\begin{table}[p]
\centering
\caption{Summary of key outputs from the conditional model component of a multilevel meta-analysis (Equation \ref{eq3}) in the \texttt{glmmTMB} R package and equivalent output in the \texttt{metafor} R package}
\label{tab:mlm-summary}
{\fontsize{8}{10}\selectfont
\begin{tabular}{
  | >{\centering\arraybackslash}m{0.14\textwidth}
  | >{\centering\arraybackslash}m{0.6\textwidth}
  | >{\centering\arraybackslash}m{0.22\textwidth}
  | >{\RaggedRight\arraybackslash}m{0.46\textwidth} |
}
\hline
 \textbf{Model output}
 & \textbf{\texttt{glmmTMB::glmmTMB}}
 & \textbf{\texttt{metafor::rma.mv}}
 & \textbf{Notes} \tabularnewline \hline

Overall mean estimate $\hat\mu$ &
\mintinline{r}{coef(summary(fit.equalto))$cond[,"Estimate"]} &
\mintinline{r}{coef(fit.metafor)} &
  \multirow{4}{=}{With moderators (fixed effects), the output becomes a vector ordered as in the model, i.e.\, intercept first followed by each moderator coefficient.}
  \tabularnewline
\hhline{|---|~|}

SE (standard error) of $\hat\mu$ &
\mintinline{r}{coef(summary(fit.equalto))$cond[,"Std. Error"]} &
\mintinline{r}{se(fit.metafor)} &
  \tabularnewline
\hhline{|---|~|}

$z$-value of $\hat\mu$ &
    \mintinline{r}{coef(summary(fit.equalto))$cond[,"z value"]} &
    \mintinline{r}{fit.metafor$zval} &
  \tabularnewline
\hhline{|---|~|}

p-value of $\hat\mu$ &
  \mintinline{r}{coef(summary(fit.equalto))$cond[,"Pr(>|z|)"]} &
  \mintinline{r}{fit.metafor$pval} &
  \tabularnewline \hline

Among-study variance estimate $\hat\tau^2$ &
  \mintinline{r}{VarCorr(fit.equalto)$cond$study[,"(Intercept)"]} &
  \mintinline{r}{fit.metafor$sigma2[1]} &
  Replace \texttt{study} with the exact name of the random-effect study-level grouping factor (the code depends on that name). If more random-effects are specified, substitute the random-effect name similarly.
  \tabularnewline \hline

Within-study variance estimate $\hat\sigma^2$ &
  \mintinline{r}{sigma(fit.equalto)^2} &
  \mintinline{r}{fit.metafor$sigma2[2]} &
  For \verb|glmmTMB|, the within-study variance estimate is the residual variance. Alternatively, \verb|exp(fit.equalto$fit$par["betadisp"])^2| yields the same value.
  \tabularnewline \hline
\end{tabular}
}
\end{table}
\vfill
\end{landscape}

\section{Application}

This section demonstrates the application of the new \texttt{equalto} implementation through two complementary parts. First, we evaluate its performance using a benchmark simulation study. After this, we present illustrative examples showing how the implementation can be applied to a range of real meta-analysis datasets, including bivariate, network, phylogenetic multilevel, and location-scale meta-analyses. 

\subsection{Simulation study}

\subsubsection*{Simulation overview}

We registered our simulation study plan in October 2025\autocite{williams_new_2025}.
The goal of our simulation study was to assess the capabilities of \texttt{glmmTMB} to fit meta-analytic models. Performance was compared to the \texttt{metafor} package, given its long-standing use and extensive validation for meta-analysis. 
We simulated data for four different comparative effect size measures: standardised mean difference (\textbf{SMD}; assuming Hedges' $g$ formulation\autocite{hedges_statistical_1985}), log response ratio (\textbf{lnRR}\autocite{hedges_metaanalysis_1999}), log odds ratio (\textbf{OR}), and log incidence ratio (\textbf{IRR}). 

We fitted the Normal–Normal random-effects model in Equations \ref{eq1}–\ref{eq2} for each of the four effect size measures using the \texttt{rma.uni} function in \texttt{metafor} and with the \texttt{equalto} structure in \texttt{glmmTMB}. 
Additionally, for the OR and IRR measures, we fitted Binomial–Normal and Poisson–Normal GLMM parameterisations to the same simulated datasets. These models were fitted using \texttt{rma.glmm} in \texttt{metafor}, which calls \texttt{lme4::glmer}\autocite{bates_fitting_2015} with a Laplacian approximation by default, and using \texttt{glmmTMB}, which also applies a Laplacian approximation.
Full details of the simulation design, including data-generating mechanisms, model specification in each function, are given in the Supplementary Material. Table~S1 summarises the simulated parameter values for each effect size measure.

\subsubsection*{Simulation results}

We summarised convergence by reporting the number and percentage of model fits with warnings (Table~S2) and errors or failed fits (Table~S3) for each model type and effect-size measure, and retained only simulation replicates in which all models fitted without warnings or errors (Table~S4). No errors occurred for the normal--normal or \texttt{glmmTMB} one-stage models, whereas one \texttt{rma.glmm\_OR} fit and 171 \texttt{rma.glmm\_IRR} fits failed.

\begin{figure}[!ht]
   \centering
   \includegraphics[width=1\linewidth]{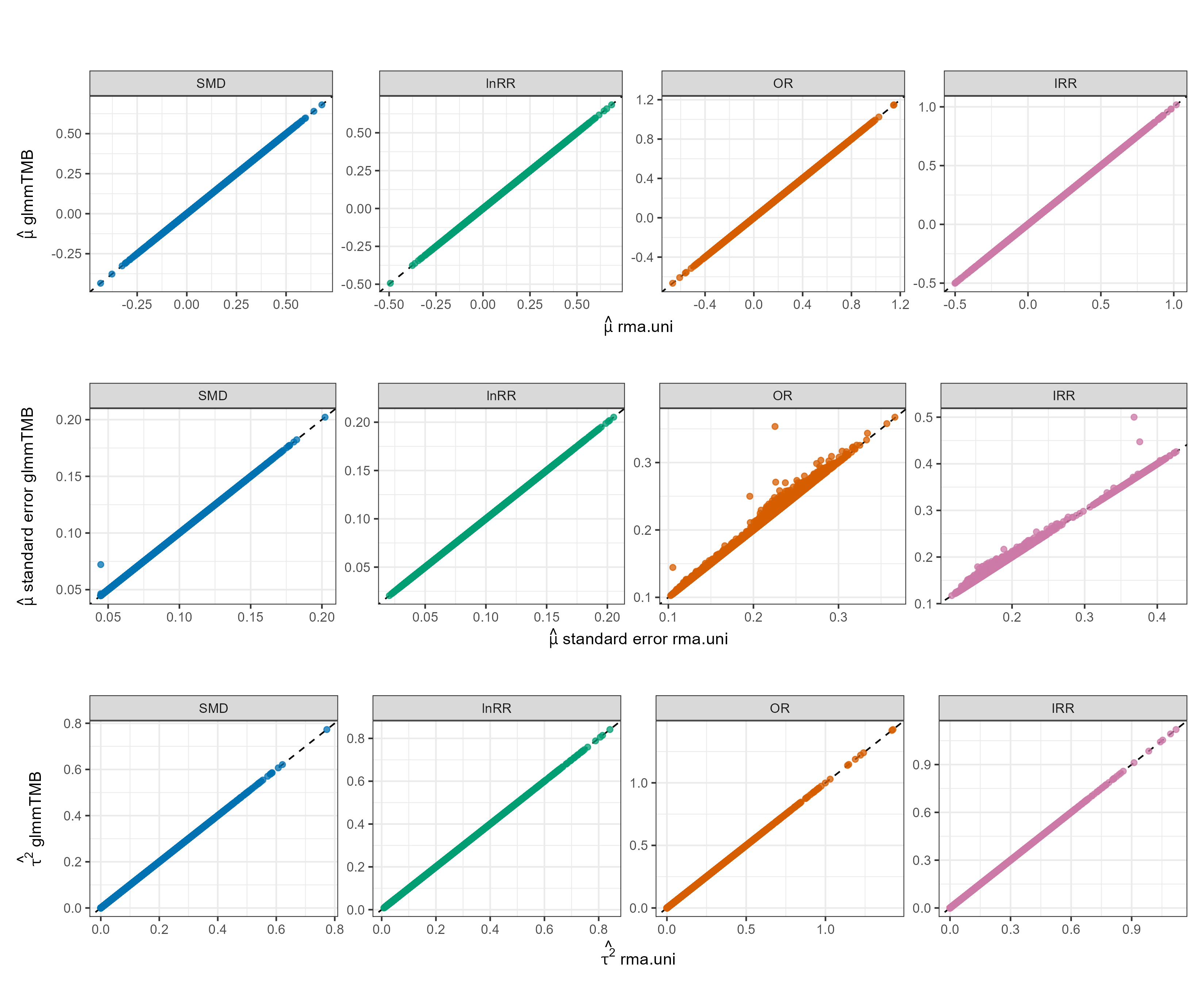}
   \caption{Agreement of Normal-Normal random-effects meta-analysis estimates between \texttt{glmmTMB::glmmTMB} and \texttt{metafor::rma.uni} across converged simulations (see Table~S4). \\
   \textbf{Top row}: overall mean estimates ($\hat{\mu}$); \textbf{middle row}: standard error of the overall mean ($\mathrm{SE}(\hat{\mu})$); \textbf{bottom row}: between-study variance estimates ($\hat{\tau}^2$) for each effect-size measure (SMD, lnRR, OR, IRR)}
   \label{fig:estimates}
\end{figure}

For the Normal–Normal models, \texttt{glmmTMB} and \texttt{rma.uni} produced identical overall mean estimates ($\hat{\mu}$) to the 6th decimal point and between-study variance estimates ($\hat{\tau}^2$) to the 5th decimal point for all effect-size measures, and mostly identical standard errors, with only a few outlying standard errors differing by approximately 0.1 for OR and IRR (Figure~\ref{fig:estimates} and Figure~S1). Additionally, RMSE distributions, coverage rates, confidence-interval widths, type I error, and power were identical for both functions (Figures~S4–S9).

For the Binomial–Normal models, \texttt{glmmTMB} and \texttt{rma.glmm} gave similar overall mean, standard-error, and between-study variance estimates (Figure~S2), with small differences only for the second \texttt{glmmTMB} specification. RMSE, coverage, confidence-interval widths, type I error, and power were similar between \texttt{glmmTMB} and \texttt{rma.glmm} (Figures~S4–S9). We also considered an alternative \texttt{glmmTMB} specification according to the model described in Supplementary Material 1.3.2 and this had slight differences compared with the other models and was generally less desirable (Figures~S4 and S9).

For the Poisson–Normal models, \texttt{glmmTMB} and \texttt{rma.glmm} produced almost identical overall mean, standard-error, and between-study variance estimates (Figure~S3), and identical RMSE distributions, coverage, confidence-interval widths, type I error, and power (Figures~S4–S9).

For the Normal–Normal models, \texttt{rma.uni} was faster than \texttt{glmmTMB} by 0.5 to 0.8 of a second on average (Figure~S10). For the GLMM specifications, \texttt{glmmTMB} was faster than \texttt{rma.glmm} by 0.5 to 0.7 of a second on average (Figure~S10).

\subsection{Illustrative examples}

To illustrate the application of the new covariance structure \mintinline{r}{equalto()} to fit meta-analytic models, we re-analyse datasets from four published meta-analyses. We provide the full R scripts and data information on the following webpage (\href{https://coraliewilliams.github.io/equalto_sim_study/webpage.html}{link}), which also serves as a tutorial for using the \texttt{glmmTMB} R package for meta-analysis. The following model results are illustrative only and should not be used to support any substantive conclusions. Their purpose is to demonstrate how the models can be specified and fitted in practice -- they are not intended to provide new empirical insights or to evaluate the questions addressed in the original studies.

\subsubsection{Bivariate meta-analysis}

Meta-analysis is widely used in medicine to pool evidence from clinical trials or observational studies. Because many outcomes are events, common effect-size measures include odds ratios, risk ratios, and risk differences. Here we use a dataset of a meta-analysis which evaluates the efficacy of the Bacillus Calmette–Guérin (BCG) vaccine against tuberculosis across 13 studies\autocite{colditz_efficacy_1994}, available in the \texttt{metadat} package. We want to fit a bivariate model to this data where the treatment groups are modelled as separate outcomes.

We reshape the data to a long format so each study contributes one row per treatment group (vaccinated “\texttt{v}” and unvaccinated/control “\texttt{c}”), defined as the \texttt{group} variable, and compute the logit-transformed proportion (log odds) as the effect-size measure that will be modelled as the outcome \texttt{yi}. 
Assuming studies and groups within studies are independent, we derive the sampling error variance-covariance matrix with sampling variances along the diagonal and zeros on the off-diagonals:

\begin{minted}{R}
VCV <- diag(dat_long$vi)
\end{minted}

To fit the bivariate model, we include treatment-specific means fixed term \texttt{group}, a random term \texttt{(0 + group|study)} to model the between-study variances for each treatment group and their covariance, and the sampling error component \texttt{equalto(0 + obs|g, VCV)}. This specification estimates the between-study variances for each treatment and their covariance, capturing how study-level effects co-vary across treatment groups. We assume no within-study variability and treat all heterogeneity as between-study variation because there is only one observed effect size per study and per treatment group in this dataset. To do this, we set \texttt{dispformula = $\sim$ 0} to fix the residual variance at (approximately) zero so that within-study variance is not modelled. 
The model is fitted in \texttt{glmmTMB} as follows:

\begin{minted}{R}
bv_fit <- glmmTMB(yi ~ group + (0 + group|study) + equalto(0 + obs|g,VCV),
                  dispformula = ~ 0,
                  REML = TRUE,
                  data = dat_long)
\end{minted}

The estimated mean logit-transformed event proportion was -4.10 for the control arm and -4.84 for the vaccinated arm, indicating a lower tuberculosis event probability in vaccinated groups (approximately 1.6\% vs 0.8\% after back-transformation). Large between-study heterogeneity was observed for both the control arm ($\sigma^2_\text{control}$ = 2.62) and the vaccinated arm ($\sigma^2_\text{vaccinated}$ = 1.55).

\bigskip

We can also fit the bivariate outcome directly as a one-stage binomial GLMM, using the numbers of events (`event`) and non-events in the treatment and control arms of each study. This corresponds to the ``Van Houwelingen bivariate'' model\autocite{jackson_comparison_2018}:

\begin{minted}{R}
# glmmTMB model one-stage approach 
bv_joint_fit <- glmmTMB(cbind(event, n-event) ~ factor(treat) 
                                                + (control+treat-1|study), 
                         family = binomial(link = "logit"), 
                         data = dat_long)
\end{minted}

\subsubsection{Network meta-analysis}

Network meta-analysis combines evidence across studies comparing multiple interventions, including treatments that have not been compared directly\autocite{salanti_indirect_2012}. We use a smoking-cessation dataset\autocite{hasselblad_metaanalysis_1998}, available in the \texttt{metadat} package, comprising 50 treatment-group observations from 24 studies comparing no contact, self-help, individual counselling, and group counselling. For each treatment group, the outcome is the number of participants who remained abstinent for 6 to 12 months (\texttt{xi}) out of the total number of participants (\texttt{ni}).

We fit a one-stage, arm-based network meta-analysis using a binomial GLMM with a logit link and use ``no contact" as the reference treatment. We use a compound-symmetry covariance structure to estimate treatment-specific between-study variances and assuming a common correlation between treatment groups from the same study.

\begin{minted}{R}
nma_fit <- glmmTMB(cbind(xi, ni-xi) ~ trt + cs(0 + trt | study),
                   family = binomial(link = "logit"),
                   data = dat)
\end{minted}

Relative to the ``no contact" approach, the odds of abstinence were estimated to be 1.69 times higher with self-help (95\% profile likelihood CI: 0.90--3.33; $p = 0.070$), 2.18 times higher with individual counselling (95\% CI: 1.52--3.27; $p < 0.001$), and 2.92 times higher with group counselling (95\% CI: 1.42--6.14; $p < 0.001$). The results therefore indicated higher abstinence odds for individual and group counselling, while the confidence interval for self-help showed no evidence of a difference from no contact.

The estimated between-study variances were $0.19$, $0.48$, $0.54$, and $0.51$ for no contact, self-help, individual counselling, and group counselling, respectively. The common within-study correlation was estimated as $\rho = 0.49$, indicating moderately correlated study-level abstinence rates across treatment arms.

\subsubsection{Phylogenetic meta-analysis}

In meta-analysis of ecological and evolutionary studies, effect sizes are often drawn from multiple species. Because species share evolutionary history, these effect size estimates are not independent\autocite{lajeunesse_metaanalysis_2009, adams_phylogenetic_2008, chamberlain_does_2012}. To account for this phylogenetic relatedness, we include a species-level random effect structured by a phylogenetic correlation derived from a phylogenetic tree and assumed model of evolution. \\

We illustrate the approach with a published meta-analysis on plant diversity effects on leaf traits of 955 effect sizes across 39 studies and 102 species\autocite{jurifelix_jurifelix_2023, felix_plant_2023}. The meta-analysis used as outcome \texttt{yi} the standardised mean difference (SMD; Hedges’ $g$ \autocite{hedges_statistical_1985}) comparing the mean trait value of a focal species in mixture versus monoculture plant environments. First, we set up the sampling error covariance matrix assuming a within-study correlation of 0.5\autocite{noble_nonindependence_2017}:

\begin{minted}{R}
V <- metafor::vcalc(vi, cluster=ACC, obs=id, data=dat, rho=0.5)
rownames(V) <- colnames(V) <- levels(ef$id)
\end{minted}

We fit a multilevel model in which experimental site (\texttt{experiment}), study ID (\texttt{ACC}), individual effect ID (\texttt{id}), and plant species (\texttt{species}) were included as random factors to account for non-independence among effect sizes. Phylogenetic correlations at the species level (\texttt{phylo.cor} matrix) were incorporated via the \texttt{propto} covariance structure.

\begin{minted}{R}
fit_phylo <- glmmTMB(yi ~  Nfixing - 1 + 
                          (1 | experiment) + (1 | ACC) +
                          (1 | species) + 
                          propto(0 + species|g, phylo.cor) + 
                          equalto(0 + id|g, V), 
                     data = ef, 
                     REML=TRUE)
\end{minted}

The estimated effects were small and not statistically significant for both conditions (N-fixing neighbours: $0.106 \pm 0.324$, $p = 0.745$; no N-fixing neighbours: $0.086 \pm 0.314$, $p = 0.785$), indicating little evidence that neighbour N-fixing status influences the standardised mean difference in leaf traits. Variance components indicated heterogeneity at the experiment ($\sigma^2_\text{experiment} = 0.204$), non-phylogenetic species ($\sigma^2_\text{non-phylo} = 0.106$), and phylogenetic species levels ($\sigma^2_\text{phylo} = 0.296$), while between-study ($\sigma^2_\text{ACC}$) variation was negligible. The estimated phylogenetic signal was $\lambda = \frac{0.296}{0.296 + 0.106} \approx 0.74$, suggesting that most species-level variation is from shared evolutionary history among species. \\

We can fit the same model with the \texttt{rma.mv} function in \texttt{metafor} as follows: 

\begin{minted}{R}
fit_phylo_rma <- rma.mv(yi, vi,
                        mods = ~ Nfixing - 1,
                        random = list( ~ 1 | experiment, ~ 1 | ACC,
                                       ~1 | species, ~1 | phylo,
                                       ~1 | id), 
                        R = list(phylo = phylo.cor),
                        data = ef,
                        method="REML") 
\end{minted}

We found that \texttt{glmmTMB} was faster than \texttt{rma.mv} for this dataset and model specification by $\sim 46$ seconds (19 vs 65 seconds respectively).

\subsubsection{Location-scale meta-analysis}

In meta-analysis, interest often lies in understanding how heterogeneity varies across studies. Standard models assume that the between-study variance $\tau^2$ is constant, an assumption that may not hold in many empirical settings. Location–scale models relax this assumption by allowing both the \emph{location} (the mean) and the \emph{scale} (the variance) to depend on moderators and random grouping terms\autocite{viechtbauer_locationscale_2022,nakagawa_locationscale_2025}, enabling tests of whether variability changes with study features (e.g.\, outcome type, design, sample size) and across subgroups.

We illustrate a location–scale model with a published meta-analysis of school-based writing-to-learn interventions on academic achievement\autocite{bangert-drowns_effects_2004}. The meta-analysis synthesised 48 studies comparing an intervention (greater emphasis on writing tasks) with a control condition. The authors derived effect-size measures of standardised mean differences (\texttt{yi}) between the two interventions, where positive values indicated better performance under the intervention. Below, the mean model includes sample size (\texttt{ni}) as a moderator, and the scale model uses \verb|dispformula| to specify that the (residual/dispersion) variance varies with study sample size (\texttt{ni}) on the log scale. The sampling error structure with variance–covariance \texttt{V} is supplied via \verb|equalto|, where we assume independence among sampling errors:

\begin{minted}{R}
V <- diag(dat$vi)
fit_ls <- glmmTMB(yi ~ ni + equalto(0 + id|g, V),
                  dispformula = ~ ni,
                  data = dat,
                  REML = TRUE)
\end{minted}

The estimated mean effect size was positive ($\mu = 0.30$, $p < 0.001$), indicating there is evidence that writing-to-learn interventions were associated with improved academic achievement. Sample size (\texttt{ni}) had a very small but statistically significant negative effect on the mean effect size ($\beta = -0.00055$, $p = 0.005$), suggesting that larger studies tended to report slightly smaller effects. In the scale model, the intercept of $-0.96$ on the log(SD) scale corresponds to a between-study heterogeneity variance of approximately $0.15$ (obtained by exponentiating $2 \times -0.96$). Further, the scale model shows no evidence that heterogeneity decreases with increasing sample size ($p = 0.071$).

This model can also be fitted in \texttt{metafor} using the \texttt{rma} function, we provide the code in the supplementary webpage (\href{https://coraliewilliams.github.io/equalto_sim_study/webpage.html}{link}). \texttt{rma} currently does not support random effects in the dispersion model (but this feature is under development), whereas \texttt{glmmTMB} does\autocite{nakagawa_locationscale_2025}.


\section{Discussion}

In this paper, we introduce the new \texttt{equalto} covariance structure in the \texttt{glmmTMB} R package, extending its existing one-stage GLMM capabilities to conventional two-stage inverse-variance meta-analysis. With this addition, \texttt{glmmTMB} provides a unified framework for fitting two-stage normal--normal models and one-stage meta-analytic GLMMs using the same model-fitting function. 
We showed that \texttt{glmmTMB} yields estimates identical to the corresponding \texttt{metafor} functions for normal--normal models and similar estimates for GLMM specifications. Below, we highlight the main advantages of \texttt{glmmTMB} for flexible meta-analytic modelling, discuss its current limitations, and outline other relevant R implementations and future directions for development.

The key advantage of using \texttt{glmmTMB} for meta-analysis is that this R package is highly flexible and can be used to fit a broad range of meta-analysis models for settings otherwise difficult to address, such as the one-stage network meta-analysis illustrated in Section 4.2.2. Further options afforded by \texttt{glmmTMB} include the ability to account for multiple fixed and random effects, support for a wide range of distributional families, a range of options for random effects with structured covariance matrices, a dedicated dispersion formula (\texttt{dispformula}), and a zero-inflation component (\texttt{ziformula}) for modelling excess zeros. A further advantage is that a common interface is used for all meta-analytic models, with the same \texttt{glmmTMB()} function and \texttt{lme4}-style formula interface\autocite{bates_fitting_2015} used for both two-stage normal--normal and one-stage models.
Compared with \texttt{rma.mv} function in \texttt{metafor} for fitting multilevel meta-analysis, \texttt{glmmTMB} has the additional advantage of allowing random effects in the dispersion model. For GLMM meta-analysis, \texttt{glmmTMB} can accommodate multiple random effects, whereas \texttt{rma.glmm} function is currently limited to a single study-level random effect and no separate dispersion formula.
In our simulations, \texttt{glmmTMB} produced fewer warnings and failed fits than the default \texttt{rma.glmm} implementation for the one-stage Poisson--normal models, although performance was similar for the binomial--normal models (Tables~S2 and~S3). By default, \texttt{rma.glmm} fits mixed models using \texttt{glmer} from the \texttt{lme4} package, but \texttt{glmmTMB} or \texttt{mixed\_model} from the \texttt{GLMMadaptive} package can alternatively be selected through its \texttt{control} argument.

However, some desirable functionalities available in \texttt{rma.mv} are not currently directly implemented in \texttt{glmmTMB}, such as the functionality to subset a given dataset for subgroup meta-analysis and options for adjusted $t$-tests for the conditional model (though these can be derived easily; see supplementary \href{https://coraliewilliams.github.io/equalto_sim_study/webpage.html#derive-adjusted-t-distribution-p-values}{webpage} for an example). 
Further, \texttt{glmmTMB} includes a \texttt{vcovHC()} function for robust sandwich covariance estimates but currently supports only the unadjusted form and no small-sample corrections or alternative estimators such as CR1--CR3 \autocite{pustejovsky_smallsample_2018}.

Regarding computational speed, in our simulation study for simple random-effects meta-analysis models (without moderators), \texttt{rma.uni} in \texttt{metafor} was slightly faster than \texttt{glmmTMB} (by about half a second). However, this may not be the case when fitting more complex models, as in our phylogenetic meta-analysis example, where we found \texttt{glmmTMB} ran approximately three times faster than \texttt{metafor}. For models with multiple random effects or with high-dimensional variance-covariance matrices, computation often becomes a bottleneck. Although \texttt{metafor}’s performance could potentially be improved by using different system libraries for matrix algebra\autocite{oancea_accelerating_2015} (e.g.\ OpenBLAS, Intel’s Math Kernel Library) or alternative optimisers, we expect \texttt{glmmTMB} to retain a speed advantage for such models (i.e.\ phylogenetic, spatial, or temporal meta-analysis).


Beyond the \texttt{glmmTMB} implementation, other general mixed-model packages also support meta-analysis, but they hold some limitations. For example, the \texttt{nlme::lme}\autocite{pinheiro_mixedeffects_2006} function, which is primarily intended for nested hierarchical structures, can fix residual variances using the \texttt{varFixed()} argument in a maximum-likelihood framework, but under REML it produces small numerical discrepancies in variance component estimates. The \texttt{mixmeta} package\autocite{sera_extended_2019} extends mixed models to meta-analysis with specialised structures (e.g.\ dose–response, longitudinal designs), but lacks some features to fit some specific covariance structures (e.g. for spatial meta-analysis). 
Furthermore, Bayesian frameworks such as \texttt{brms}\autocite{burkner_brms_2017} provide similarly flexible options for complex meta-analysis (e.g.\ phylogenetic or location-scale meta-analysis), and can also accommodate different sampling error covariance structures. However, they rely on Monte Carlo Markov Chain sampling, which is typically more computationally demanding for high-dimensional models than the maximum-likelihood optimisation used by \texttt{glmmTMB}.

\bigskip
By adding \texttt{equalto} to \texttt{glmmTMB}, we extend an already flexible GLMM framework into a broader meta-analytical toolkit that can accommodate both conventional two-stage inverse-variance models and one-stage meta-analytic GLMMs.
This unified framework, within the same \texttt{glmmTMB()} function, can support diverse distributional families, complex dependence and multivariate structures, dispersion models, and zero-inflation components, and both one- and two-stage individual participant data meta-analysis (IPD-MA)\autocite{burke_metaanalysis_2017}.

Because we do not intend to replace existing meta-analysis software for post-hoc analysis and visualisation, we instead emphasise compatibility with existing tools, as demonstrated by the orchard plots\autocite{nakagawa_orchard_2023} presented on the accompanying webpage, while future work could extend this compatibility more broadly. 
Additionally, future work could involve investigating different estimation methods for the GLMM meta-analysis specifications beyond Laplace approximation, which we used in our simulations, for example Gauss-Hermite quadrature and penalized quasilikelihood\autocite{bolker_generalized_2009}, alongside simulation studies reflecting more realistic conditions, including varying numbers of effect sizes per study, smaller number of studies and different dependence structures.


\newpage
\section*{}

\begin{Backmatter}

\paragraph{Acknowledgments}
We thank the developers of the \texttt{glmmTMB} and \texttt{TMB} R packages for their work in making these packages open-source and widely available. Thank you to Michael McCarthy and James Pustejovsky for their helpful comments on an earlier version of this manuscript. 

\paragraph{Funding Statement}
This study was supported by an Australian Research Council, ARC, Discovery Grant (DP230101248) and CERC (Canada Excellence Research Chair; CERC-2022-00074) awarded to SN. 

\paragraph{Competing Interests}
CW and MM are contributors to the \texttt{glmmTMB} R package, and MB and BB are authors and maintainers.
WV is the author and maintainer of the \texttt{metafor} R package and contributor to the \texttt{metadat} R package. 

\paragraph{Data Availability Statement}
Data and code for the simulations and illustrative examples can be found at the following link: \url{https://doi.org/10.5281/zenodo.22742805}.

\paragraph{Author Contributions}
Conceptualisation: CW, MM, DW, SN.
Data Curation: CW.
Formal Analysis: CW.
Funding Acquisition: CW, SN.
Investigation: CW.
Methodology: CW, MM, YY, SN.
Project Administration: CW.
Software: CW, MM, MB, BB.
Supervision: SN, DW.
Validation: WV, MB, BB.
Visualization: CW, AM. 
Writing – Original Draft Preparation: CW.
Writing – Review \& Editing: All authors. 

\subsubsection*{}
The authors wrote the original manuscript and used OpenAI ChatGPT (GPT-5) solely to improve clarity and readability in some parts of the text; all methods, analyses, and conclusions were developed by the authors.



\newpage
\printbibliography

\end{Backmatter}

\end{document}